\documentclass[11pt]{article}

\usepackage[T1]{fontenc}
\usepackage{newtxtext,newtxmath}
\usepackage[margin=1in]{geometry}
\usepackage{graphicx}
\usepackage{amsmath}
\usepackage{booktabs}
\usepackage{array}
\usepackage{tabularx}
\usepackage[hidelinks]{hyperref}
\usepackage{url}
\makeatletter
\renewcommand\normalsize{%
  \@setfontsize\normalsize{11}{13.6}%
  \abovedisplayskip 11\p@ \@plus3\p@ \@minus6\p@
  \abovedisplayshortskip \z@ \@plus3\p@
  \belowdisplayshortskip 6.5\p@ \@plus3.5\p@ \@minus3\p@
  \belowdisplayskip \abovedisplayskip
  \let\@listi\@listI}
\makeatother
\normalsize
\providecommand{\doi}[1]{\url{https://doi.org/#1}}

\newcolumntype{Y}{>{\raggedright\arraybackslash}X}
\title{\fontsize{18}{22}\selectfont\textbf{Measuring Student Self-Assessment against Viva-Demonstrated Mastery in a Large First-Year Programming Course}}
\author{%
\large Sakshi Sharma$^{1,*}$, Pavani Ayinampudi$^{2}$, Aditya B.M.V.$^{2}$, Jinal Gupta$^{2}$,\\
\large Prakash Hegade$^{2}$, Rohit Sharma$^{1}$, Meenakshi V$^{1}$, S.R.S. Iyengar$^{1}$\\[8pt]
\normalsize $^{1}$Indian Institute of Technology Ropar, Rupnagar, Punjab, India\\
\small\texttt{\{sakshi.23csz0006, rohit.24csz0014,}\\
\small\texttt{meenakshi.19csz0013, sudarshan\}@iitrpr.ac.in}\\[4pt]
\normalsize $^{2}$ANNAM.AI, Indian Institute of Technology Ropar, Rupnagar, Punjab, India\\
\small\texttt{\{pavania.harvard2025, adityabmv, jinalbirla, prakash.hegade\}@gmail.com}\\[6pt]
\small $^{*}$Corresponding author: \texttt{sakshi.23csz0006@iitrpr.ac.in}}
\date{}

\begin{document}
\maketitle

\begin{abstract}
Mastery-based education increasingly places the reporting of learning progress in students' hands, who record task completion on learning dashboards. The usefulness of such self-reports depends on how closely reported mastery corresponds to demonstrated competence. Most evidence on student self-assessment compares an overall self-rating with an overall examination score and therefore provides limited evidence about which tasks or which students account for the mismatch. This study examines first-year students' self-assessment against viva-demonstrated mastery at the level of individual tasks across a ladder of sixty programming tasks. The study draws on a large first-year programming course taught in 2023, involving 203 students and 12 examiners, in which every reported task was verified through an oral viva. Because a task entered the Viva only after it was reported, the design is one-sided and captures over-estimation but not under-estimation. Of 11,093 reported tasks, 10,885 (98.1\%) were demonstrated, indicating a high degree of correspondence between self-report and demonstrated mastery. The remaining 208 overestimations were not evenly distributed. A small number of students accounted for most of the errors, and they occurred mainly on difficult tasks near the end of the task ladder rather than on higher-point tasks. This task-level analysis shows that high overall self-assessment accuracy can coexist with specific areas where reported and demonstrated mastery diverge. It also provides a practical basis for directing additional verification and formative feedback toward students and tasks where such divergence is more likely.
\end{abstract}


\section{Introduction}
The skill set and deliverables of every course are measured against its designed learning outcomes. Achievement of those outcomes is assessed at the end of the course, while student progress is tracked periodically through assessments. A single examination at the end of a course provides evidence of overall achievement, but it comes too late to establish whether each competency has been mastered during the learning process. Periodic tracking provides students with limited insight into what they have and have not mastered, even though such understanding is essential for interpreting and effectively using feedback. Mastery-based education addresses this by requiring students to demonstrate individual competencies before progressing to subsequent ones (Bloom, 1968; Harden, 2002). Such mastery, therefore, requires that learning outcomes be represented through smaller tasks rather than a single large problem, since a novice's understanding is itself assembled from pieces, as described in the Knowledge-in-Pieces (KiP) framework (diSessa, 1993).

As class sizes grow, recording mastery for individual tasks cannot rely entirely on instructor judgment. Students can provide the first-level record themselves by reporting the tasks they believe they have mastered, with a dashboard that records and displays their progress (Boud et al., 1999; Bodily \& Verbert, 2017). This makes task-level progress visible to the learner and reduces the need for an instructor to manually record every mastery claim. However, a student's report is a judgment rather than direct evidence of demonstrated competence, and students may report mastery for a task they cannot subsequently demonstrate (Andrade, 2019; Falchikov \& Boud, 1989). The reported mastery, therefore, needs to be compared with an independent demonstration to determine how accurately the student's record reflects actual performance.

The correspondence between a learner's judgment and their demonstrated performance is commonly studied as calibration (Dunlosky \& Rawson, 2012; Nelson \& Narens, 1990). In this study, calibration is used in a specific sense: the extent to which a student's reported mastery corresponds to what the student can demonstrate. Existing self-assessment research has largely examined this correspondence by comparing an overall self-rating with an overall examination or instructor score (Falchikov \& Boud, 1989; Zell \& Krizan, 2014). Such comparisons can reveal whether individual students tend to over- or underestimate their overall performance, but they provide little information about the specific competencies or tasks in which the mismatch occurs. This distinction matters in mastery-based learning because a student may judge some competencies accurately while misjudging others. A task-level comparison is therefore needed to locate where reported mastery diverges from demonstrated mastery.

This study addresses this gap by comparing students' task-level self-reports of programming mastery with their performance in an oral viva. The study examines the correspondence between reported and demonstrated mastery at the cohort, student, and task levels, with particular attention to instances where reported mastery diverges from demonstrated mastery.

\section{Literature Survey}
\label{sec:bg}
Student self-assessment becomes important when learning is organised around the achievement of individual competencies rather than a single end-of-course score. A learner who can recognise what has and has not been mastered can use that information to decide what to practise next. However, the literature shows that students' judgements do not always correspond to their demonstrated performance. Falchikov and Boud's (1989) meta-analysis found systematic differences between student self-ratings and instructor judgements, while later reviews show that the correspondence varies considerably across learners, tasks, and assessment conditions (Andrade, 2019). This variability is important for mastery-based learning because a progress system that depends on students reporting their own mastery is useful only to the extent that those reports provide a dependable signal. Self-assessment accuracy is therefore not something that can simply be assumed; it needs to be examined in the context in which the judgement is made.

The question of why students' judgements diverge from their performance is closely related to metacognitive monitoring, or the ability to monitor one's own knowledge and performance (Flavell, 1979). Research has shown that monitoring can be particularly difficult when learners have limited knowledge of the material, with less knowledgeable learners sometimes having less insight into their own gaps (Ehrlinger et al., 2008; Kruger \& Dunning, 1999). At the same time, the relationship between perceived and actual competence is not uniformly weak: a metasynthesis by Zell and Krizan (2014) found a moderate correspondence overall, while studies such as Sadler and Good (2006) show that self-assessment can become highly accurate when students are given explicit criteria against which to judge their work. These findings suggest that the important question is not simply whether students are accurate, but under what task and assessment conditions their judgements become less accurate. This shifts the problem from measuring an overall level of self-assessment accuracy to locating the specific competencies at which self-judgement breaks down.

The grain at which self-assessment has typically been studied makes this difficult. Much of the literature compares one overall self-rating with one examination or instructor score (Falchikov \& Boud, 1989; Zell \& Krizan, 2014). Such comparisons can establish that a student tends to over- or under-estimate performance, but they cannot identify the particular learning tasks responsible for the mismatch. This distinction matters in a mastery-based setting because mastery is built from individual competencies: a student may judge several simple competencies accurately while misjudging a more demanding one. The knowledge-in-pieces perspective provides one explanation for why such task-level differences might arise. Novices may initially organise knowledge as loosely connected, context-dependent pieces that work in particular situations but have not yet been integrated into a coherent understanding (diSessa, 1988, 1993). A student may therefore recognise enough of a solution to believe that a task has been mastered while being unable to integrate the necessary knowledge when asked to demonstrate it. A task-level comparison between reported mastery and demonstrated performance can make this point of divergence observable.

This issue becomes particularly relevant when self-assessment is embedded in mastery-based and outcome-based learning environments. Mastery learning moves assessment toward the demonstration of individual competencies rather than relying solely on a final examination (Bloom, 1968; Harden, 2002). In large courses, dashboards and gamified environments can make progress visible through task completion, points, rankings, and other indicators (Deterding et al., 2011; Bodily \& Verbert, 2017). Such systems improve the visibility and scalability of progress tracking, but they also introduce a dependency: when students themselves mark tasks as completed, the quality of the dashboard's mastery signal depends on the accuracy of those reports. The literature therefore provides a strong basis for using self-assessment in learning environments, but offers less guidance on how to verify the signal when mastery is recorded task by task.

First-year programming provides a particularly useful setting for examining this problem. Programming requires students to coordinate conceptual knowledge with the ability to produce and explain working code, and introductory programming courses continue to present substantial learning difficulties (Bennedsen \& Caspersen, 2019; Watson \& Li, 2014). Automated assessment can efficiently determine whether submitted programs produce expected outputs and can support large-scale programming instruction (Keuning et al., 2018), but successful execution alone does not necessarily establish that a student understands the program they have produced. Oral assessment can complement such evidence by requiring students to explain and defend their work (Huxham et al., 2012; Joughin, 1998). A task-level comparison between a student's self-report and an oral demonstration therefore provides an opportunity to examine whether a dashboard claim of mastery corresponds to demonstrable competence.

The literature consequently leaves a specific gap at the intersection of these strands. Research on self-assessment and metacognitive monitoring establishes that students' judgements can diverge from demonstrated performance, but predominantly measures that divergence at an aggregate level. Research on mastery-based dashboards shows the value of making progress visible, but the self-reported mastery signal is rarely examined task by task against an independent demonstration of competence. What remains unclear is where self-report remains dependable and where it begins to diverge from demonstrated mastery within a sequence of progressively demanding tasks. This study addresses that gap by pairing each student's self-report with an oral viva outcome for each task across a sixty-task programming ladder. This task-level design allows the mismatch to be examined both across students and across tasks, providing a finer account of self-assessment accuracy than a single overall self-rating compared with a single examination score.

\section{Methodology and Methods}
\label{sec:method}

\subsection{The design}
\begin{figure}[t]
\centering
\includegraphics[width=\textwidth]{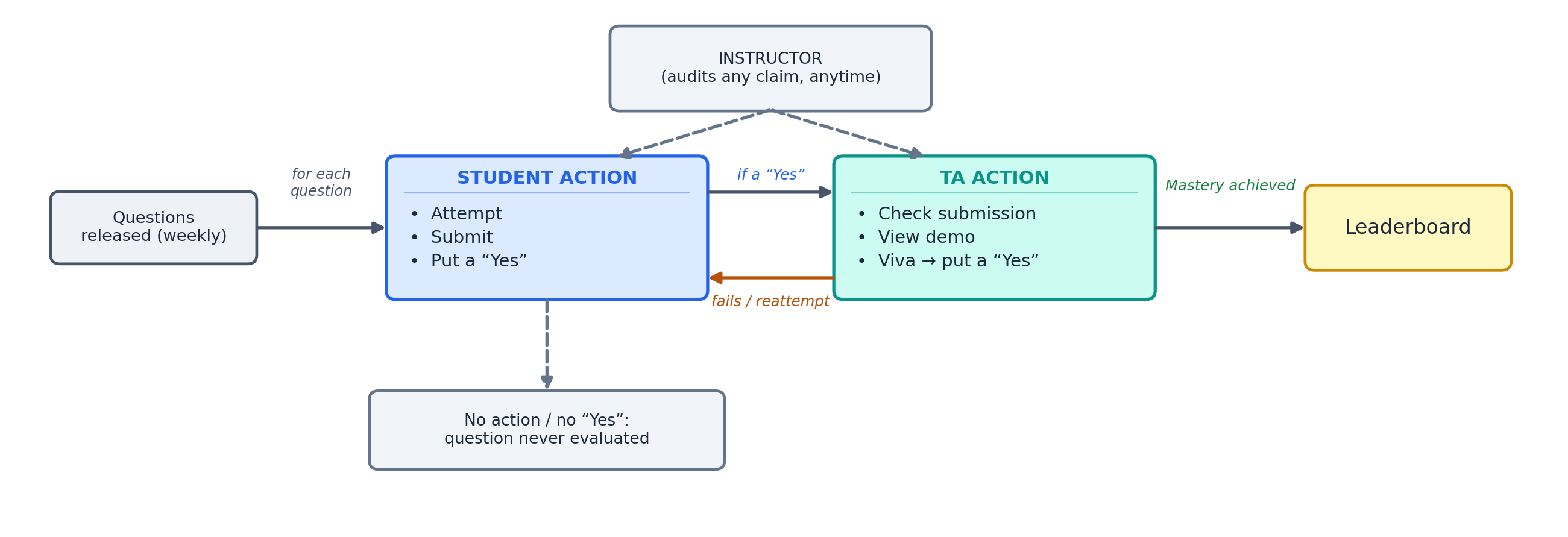}
\caption{The self-report and evaluation model.}
\label{fig:design}
\end{figure}

The study compares, for every task, what a student reported against what the same student could demonstrate at an independent oral viva. The two records are collected separately: a self-report, entered by the student on the dashboard, and a viva verification, recorded by a teaching assistant who examined the reported task. Verification is gated on the self-report. A teaching assistant examined a task only after the student had reported it, so a task that a student never reported was never examined (Fig.~\ref{fig:design}).

This gating mechanism makes the design one-sided. Students can overestimate their ability by reporting a task that they are unable to demonstrate, but they cannot underestimate their ability because tasks that are not reported are not assessed in the viva.

The viva was not limited to a single attempt. Students who reported a task but were unable to demonstrate it could reattempt the task as many times as necessary until they passed. The records, however, retain only the final outcome for each task and not the number of attempts. Thus, a task recorded as ``not demonstrated'' represents a task that the student reported but had not passed when the task closed, despite having the opportunity to reattempt it. Overestimation therefore represents a persistent gap between a student's reported ability and demonstrated ability, rather than a single failed attempt.

Calibration is measured using the overestimation rate:
\begin{center}
\textit{P}(not demonstrated $\mid$ reported)
\end{center}
This represents the probability that a reported task is ultimately not demonstrated. The measure is examined at three levels: the cohort, the individual student, and the individual task. The three research questions that follow correspond to these levels.

Because the design records only reported tasks and does not collect confidence ratings, the term calibration is used here in a specific and limited sense. It refers to the accuracy of students' reported abilities and does not capture under-reporting or the broader relationship between confidence and accuracy examined in the metacognition literature.

\subsection{Research questions}
The design permits the mismatch between reported and demonstrated mastery to be examined at three levels: across the cohort, across students, and across tasks. The research questions follow these three levels.

\smallskip
\noindent\textbf{RQ.} How well does first-year engineering students' self-assessment of programming-task completion align with their viva-demonstrated mastery, and how is the overestimation distributed across students and tasks?
\begin{itemize}
\item \textbf{SQ1 (the cohort).} How closely does self-report match viva demonstration across the cohort as a whole?
\item \textbf{SQ2 (across students).} How is overestimation distributed across students?
\item \textbf{SQ3 (across tasks).} How does overestimation vary with a task's position on the ladder and with its point value?
\end{itemize}

\subsection{Checking the criterion}
Calibration is meaningful only to the extent that viva-demonstrated mastery provides a sound criterion against which student reports can be compared. The criterion is therefore checked before the calibration analyses. Two checks are applied, addressing the behaviour of the task ladder and the consistency of the viva assessment.

The first check asks whether the sixty-task ladder behaves as a coherent measurement scale. Task-level performance is examined using item facility and corrected item-total point-biserial correlations, while internal consistency for the dichotomous task outcomes is assessed using KR-20 (Kuder \& Richardson, 1937). A two-parameter logistic item response model is also fitted to estimate the difficulty and discrimination of each task (de Ayala, 2009; Embretson \& Reise, 2000).

The second check asks whether the viva criterion is applied consistently across the twelve examiners. Per-examiner confirmation rates are compared, and between-examiner variation is summarised using a beta-binomial intraclass correlation (Koo \& Li, 2016; Shrout \& Fleiss, 1979). One structural limitation is present in this comparison: each examiner assessed their own students, so examiner and cohort are not independent. The examiner-consistency analysis is therefore interpreted as a check on variation in the observed criterion, rather than as a fully crossed estimate of examiner reliability.

Together, these checks establish the limits within which the subsequent calibration results can be interpreted.

\subsection{Analytical methods}
All analyses are scripted, and the analysis pipeline first reproduces the course's canonical descriptive statistics before any inference is drawn (Section~\ref{sec:repro}). The subsequent analyses follow the three levels defined by the research questions.

For SQ1, the cohort-level question, we compute the overall agreement between self-report and viva demonstration, together with its complement, the overestimation rate. Because tasks carry different point values, points-weighted agreement is also reported. These quantities describe how closely reported mastery corresponds to demonstrated mastery across the cohort as a whole.

For SQ2, the student-level question, we count the number of over-estimations made by each student and examine how unevenly these errors are distributed using the Gini coefficient and Lorenz curve. Because students report different numbers of tasks, the opportunity to incur an overestimation is not equal across students. We therefore benchmark the observed concentration against an exposure-adjusted null model that holds each task's difficulty fixed and randomises only which students who reported a task fail to demonstrate it. Confidence intervals are obtained by bootstrapping students with 5,000 resamples using a fixed seed.

For SQ3, the task-level question, we compute the overestimation rate for each task and examine how it varies with the task's position on the ladder. Because the observed errors form a near-step distribution, the primary comparison contrasts the early tasks with the harder late tasks using the Mann--Whitney test. A permutation test on the association between overestimation and task position provides a complementary analysis. Spearman correlation is reported as a secondary monotone summary because ladder position is confounded with both task difficulty and curricular time. Task point value is examined separately to determine whether overestimation is associated with the stakes attached to a task rather than only with its difficulty.

The study involves several correlation-based analyses, but no formal multiplicity correction is applied. The main interpretation therefore rests on the magnitude and uncertainty of the observed effects, reported with confidence intervals, rather than on borderline significance alone.

\section{Data Collection and Analysis}
\label{sec:data}

\subsection{The course and the task ladder}
The data come from a large first-year programming course taught in 2023. Learning was organised as a ladder of sixty sequenced tasks, labelled O1 to O60 by ascending position, running from routine exercises to progressively harder programming problems. The first task requires students to print a line of text, while the last implements RSA encryption. Students worked through the tasks at their own pace as they were released in weekly batches and recorded the tasks they had completed on a spreadsheet-based dashboard.

Each reported task was then verified through an oral viva. To claim a task, a student submitted their code and screenshots of its output to the course's online classroom and demonstrated the same code to a teaching assistant. Demonstrations were conducted mainly during scheduled Thursday and Friday laboratory sessions and otherwise by appointment, with an assistant spending approximately three to five minutes per task. All assistants followed a common rubric: the submission had to be present, the code had to run and produce the correct output, and the student had to answer a question posed by the assistant about the same code. Code that the student could not explain was not accepted. As noted in Section~\ref{sec:method}, a student who could not demonstrate a reported task could re-attempt its viva until passing, and only the final outcome for each task was retained. The instructor also periodically re-examined randomly selected claimed tasks using the same rubric as the assistants; these audits almost always upheld the original decision.

The tasks also carried point values that contributed to the course grade, while a public leaderboard displayed each student's standing. The resulting activity was therefore both points-based and gamified. Point values were assigned by the instructor rather than determined by difficulty alone. They ranged from 1 to 25 and summed to 162 across the sixty tasks. The two highest-valued tasks were O60, the hardest coding problem, and O45, a single non-coding reading-and-summary task included as an accessible opportunity for a struggling student to earn points. A student's mastery score is defined as the points-weighted percentage of tasks demonstrated in the viva.

The analytic sample consists of 203 students who were assigned to a branch and appear in both the self-report and viva records. The students came from five engineering branches: Computer Science (84), Mathematics and Computing (35), Chemical (33), Civil (31), and Artificial Intelligence (20). Twelve teaching assistants conducted the vivas, with each responsible for approximately seventeen students. For each student and each of the sixty tasks, the dashboard therefore provides two aligned binary records: whether the student reported the task and whether the viva confirmed it. This produces 12,180 student-task cells, of which 11,093 correspond to tasks reported by the student.

\subsection{The paired grid and its one-sided structure}
\label{sec:grid}
The two records form a $203 \times 60$ student-task grid in which each cell takes one of three observed states: reported and demonstrated, reported but not demonstrated, or not reported (Table~\ref{tab:grid}). As established by the study design in Section~\ref{sec:method}, a task could reach the viva only after it had been reported. The fourth possible combination, demonstrated but not reported, therefore cannot occur in the data.

\begin{table}[t]
\caption{Outcomes for the 12{,}180 student-task cells in the $203 \times 60$ grid.}
\label{tab:grid}
\footnotesize
\centering
\begin{tabular}{@{}lrr@{}}
\hline
\textbf{Cell type} & \textbf{Count} & \textbf{\% of cells} \\ \hline
Reported and demonstrated & 10{,}885 & 89.4\% \\
Reported, not demonstrated & 208 & 1.7\% \\
Not reported & 1{,}087 & 8.9\% \\ \hline
Total & 12{,}180 & 100.0\% \\ \hline
\end{tabular}
\par\smallskip
{\footnotesize Over-estimation rate: $208/11{,}093 = 1.9\%$ of reported tasks.}
\end{table}

This structure determines what can be learned from the paired records. The only observable disagreement is a task that a student reported but could not demonstrate, which is the overestimation defined in Section~\ref{sec:method}. Under-estimation cannot be observed because an unreported task is never assessed. The paired data therefore do not support a symmetric comparison of two independently observed binary records. Statistics that depend on both directions of disagreement, such as McNemar's test and Cohen's kappa, are consequently not informative here. The analysis instead uses the one-sided over-estimation rate, \textit{P}(not demonstrated $\mid$ reported), examined at the cohort, student, and task levels defined by the research questions.

\subsection{Reproducibility, data, and ethics}
\label{sec:repro}
The analysis pipeline is fully scripted in Python 3.9 using NumPy, SciPy, statsmodels, and girth. A single driver runs the complete pipeline from data loading through the reported statistics and figures. The loader first reproduces the course's canonical descriptive statistics before any inferential analysis is performed. Randomised procedures, including the exposure-adjusted null model and the 5,000-sample bootstrap confidence intervals, use a fixed seed so that the reported results are deterministic.

Reproducibility in this study therefore, concerns the analysis pipeline rather than the portability of the course itself. The task set, its point weights, and the ordering of the sixty-task ladder are specific to this course and would differ in another instructional setting. The de-identified data and analysis code will be released on acceptance to permit reproduction of the reported analyses.

Students were informed that their course records would be used for research. The study uses de-identified secondary course records, with students and teaching assistants pseudonymised and all results reported in aggregate.

\section{Results and Discussion}
\label{sec:results}
Table~\ref{tab:rqs} summarises the research questions, the methods used to address them, and the headline results. We first establish whether viva-demonstrated mastery provides a sufficiently sound criterion for comparison. We then examine self-report at the cohort, student, and task levels, following SQ1--SQ3 in turn. The results show a consistent pattern: self-report is highly accurate in aggregate, but the small amount of over-estimation that remains is concentrated among a minority of students and at the hardest end of the task ladder.

\begin{table}[t]
\caption{Research questions, methods, and headline results.}
\label{tab:rqs}
\footnotesize
\begin{tabularx}{\textwidth}{@{}c Y Y Y@{}}
\hline
\textbf{RQ} & \textbf{Question} & \textbf{Method} & \textbf{Key result} \\ \hline
SQ1 & How accurate is self-assessment overall? & agreement and over-estimation rate & 98.1\% of
reported tasks demonstrated; 1.9\% over-estimation \\ \hline
SQ2 & How is over-estimation distributed across students? & Gini and Lorenz;
exposure-adjusted null & Concentrated in a minority; 78\% perfectly calibrated; Gini 0.85 [0.81, 0.89] versus an
exposure-adjusted null of 0.57, and still 0.95 among the 89 who reported all nine hard tasks \\ \hline
SQ3 & How does over-estimation vary across tasks? & Early-vs-late contrast;
report-frequency partial & Confined to the hard late coding tasks O52 to O60 (0.0\% vs
17.6\%, permutation $p < 0.001$); none on the first 51, and unrelated to point value \\ \hline
\end{tabularx}
\end{table}

\subsection{The viva criterion and its limits}
\label{sec:criterion}
Before calibrating self-report against the viva, we first examine whether viva-demonstrated mastery provides a sufficiently sound criterion. The sixty-task ladder is internally reliable, with KR-20 = 0.94 across all tasks and 0.95 across the 45 tasks that show response variance. The corrected item-total point-biserial correlations are also healthy, with a median of 0.43 and a range of 0.18 to 0.83; only two tasks fall below 0.2. These statistics, however, need to be read in light of the way the criterion is constructed: a task that a student never reported is recorded as not demonstrated. The resulting reliability therefore reflects not only the viva's ability to distinguish mastery but also how far students progressed along the ladder.

The task ladder also shows a strong ceiling effect. The mean task facility is 0.89, and 15 of the 60 tasks were demonstrated by every student. The two-parameter item response model gives a mean difficulty of \textit{b} = $-1.76$, indicating that most of the ladder's discriminating information is concentrated in the hardest, latest tasks. The criterion therefore provides substantially more information at the difficult end of the ladder than at the easy end.

The examiner-level results provide weaker evidence of consistency. Across the twelve examiners, confirmation rates range from 93.8\% to 100\%, while the between-examiner intraclass correlation is small (ICC = 0.055, 95\% CI [0.02, 0.11]). Because examiners assessed their own students, the examiner and cohort are entangled, so this value cannot distinguish examiner leniency from differences in student ability. It should therefore be treated as a weak assurance of consistency rather than as a strong estimate of examiner reliability. If some examiners were systematically more lenient, the observed 98.1\% agreement between reported and demonstrated tasks would represent an upper bound on agreement under a common standard.

The instructor's periodic re-examination of randomly selected claimed tasks under the same rubric provides an additional check. These audits almost always upheld the assistants' decisions, giving some independent assurance that the rubric was applied faithfully, although the audits cannot rule out the possibility that the same standard was applied consistently but differed from the intended standard.

Taken together, the evidence supports treating viva-demonstrated mastery as an adequate but imperfect criterion. Its measurement information is concentrated in the harder tasks, and examiner consistency is only partly identifiable because examiner and cohort are confounded. The calibration results are therefore interpreted within these limits.

\subsection{SQ1: Self-assessment is highly accurate in aggregate}
With the viva treated as the criterion within the limits established above, SQ1 asks how closely self-report matches demonstrated mastery across the cohort. Of the 11,093 tasks that students reported, 10,885, or 98.1\%, were demonstrated at the viva. The points-weighted agreement is almost identical at 98.0\%, leaving 208 reported tasks, or 1.9\%, as over-estimations. At the cohort level, therefore, students' reports of what they had completed corresponded closely to what they could demonstrate.

This high aggregate agreement, however, does not represent two-directional calibration. Verification was gated on self-report, so the analysis can detect over-claiming but cannot observe under-reporting. In addition, students who progressed less far along the ladder contributed fewer opportunities to report the hardest tasks. The 98.1\% figure should therefore be read as the rate at which reported mastery was confirmed, rather than as a general measure of calibration across all possible task judgements.

The aggregate result is nevertheless important because it establishes the overall pattern against which the remaining analyses are read. A dashboard designer might take 98.1\% agreement as sufficient reason to trust student self-report. The student-level and task-level analyses show why that conclusion would be premature.

\subsection{SQ2: Over-estimation is concentrated in a minority of students}
\begin{figure}[t]
\centering
\includegraphics[width=0.95\textwidth]{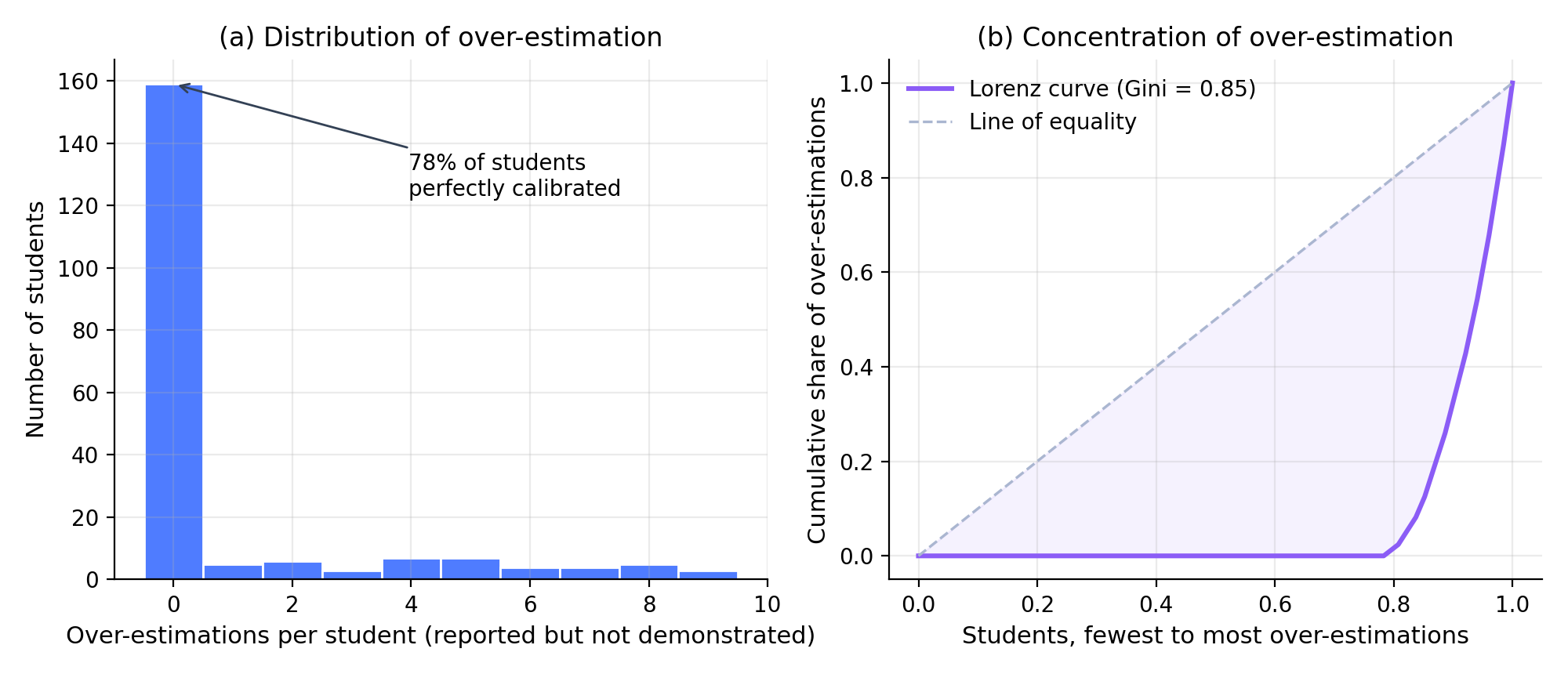}
\caption{Concentration of over-estimation across students: (a) over-estimations per student, 78\% at zero; (b) Lorenz curve, Gini 0.85.}
\label{fig:conc}
\end{figure}

The high aggregate agreement hides substantial heterogeneity between students. Of the 203 students, 159 (78\%) were perfectly calibrated in the observed data, demonstrating every task they reported. The remaining 44 students (22\%) accounted for all 208 over-estimations, averaging 4.7 over-estimations each and reaching a maximum of nine. The distribution is therefore highly unequal: the Gini coefficient is 0.85 (95\% CI [0.81, 0.89]), with the bootstrap resampling students, and the least-calibrated 10\% of students account for 67\% of all over-estimation (Fig.~\ref{fig:conc}).

The concentration could partly arise because students have different numbers of reported tasks and therefore different opportunities to over-estimate. We therefore compare the observed concentration with an exposure-adjusted null that holds each task's difficulty fixed and randomises only which students who reported a task fail to demonstrate it. The observed Gini of 0.85 is substantially higher than the null value of 0.57 (95\% CI [0.54, 0.61]; $p < .001$). The observed concentration is therefore greater than would be expected from task difficulty and unequal exposure alone under the specified null model.

The same pattern remains when exposure is held more closely constant. Among the 89 students who reported all nine hard late tasks, O52 to O60, 83 (93\%) were perfectly calibrated. The over-estimation among these students is even more concentrated, with a Gini coefficient of 0.95. Three students reported all nine of these tasks but demonstrated none of them. Thus, even among students exposed to the same hard-task tail, most demonstrated what they reported while a small number accounted for a disproportionate share of the errors.

The distribution also differs across branches. Civil students, who make up 15\% of the cohort, account for 47\% of all over-estimation (97 of 208), and 58\% of Civil students over-estimated at least once. In contrast, no Artificial Intelligence student over-estimated and fewer than 1\% of reported Computer Science tasks were over-estimations. This branch pattern should be interpreted cautiously because most teaching assistants examined students within a single branch, making examiner and branch effects inseparable, and because struggling groups received additional support during the term (Section~\ref{sec:limits}). The data therefore identify where over-estimation is concentrated, but they do not establish a branch-level cause.

At the student level, the main result is consequently not a gradual increase in error across the cohort. Most students report tasks they can demonstrate, while a small minority account for a disproportionate share of the observed over-estimation.

\subsection{SQ3: Over-estimation concentrates on the hardest tasks}
\begin{figure}[t]
\centering
\includegraphics[width=0.95\textwidth]{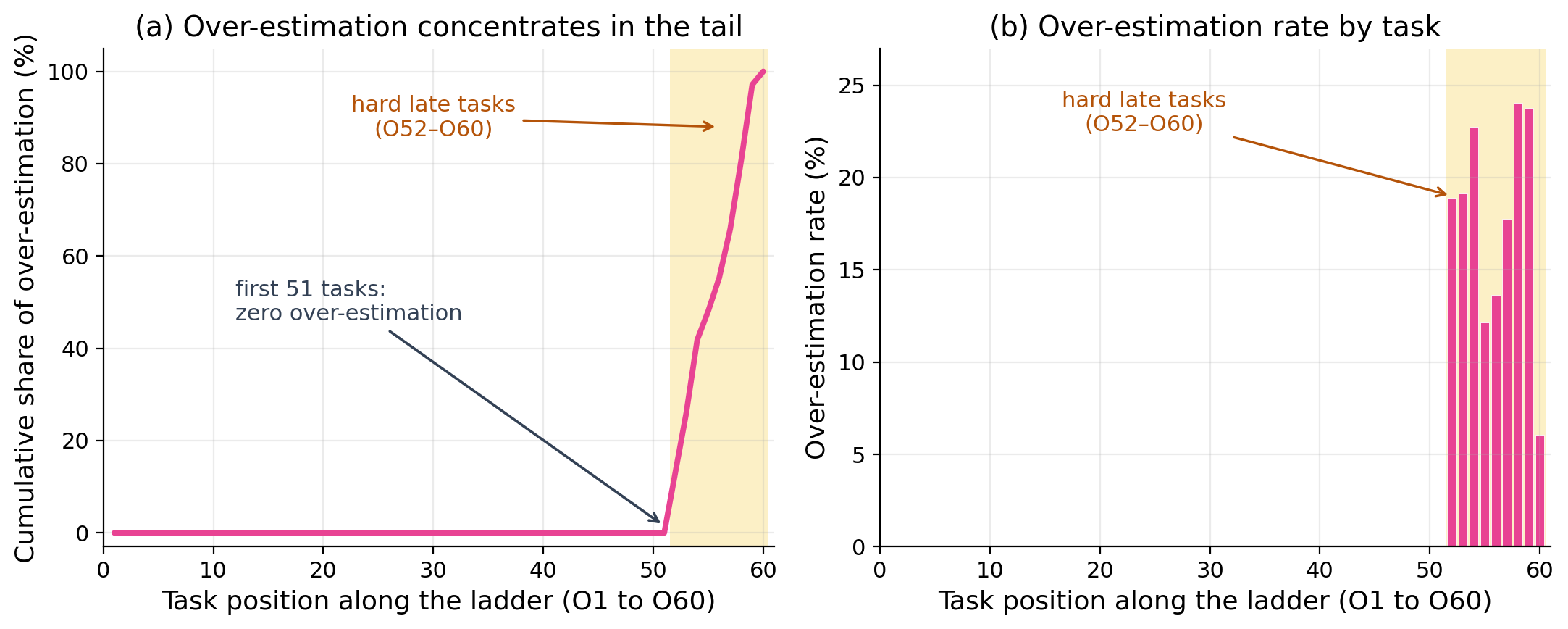}
\caption{Over-estimation along the 60-task ladder: (a) cumulative share; (b) per-task rate. All observed over-estimation occurs on the hard late tasks O52 to O60.}
\label{fig:drift}
\end{figure}

The task-level analysis shows where the over-estimation occurs. None of the 208 over-estimations occurs before task O52. The first 51 tasks therefore have an over-estimation rate of exactly zero, while tasks O52 to O60 account for all observed over-estimation, with a mean rate of 17.6\% (Fig.~\ref{fig:drift}). The O52 boundary was identified from the observed data rather than specified in advance, so it should be interpreted descriptively as the point at which the errors begin, not as a prespecified threshold.

The resulting pattern is a step rather than a gradual increase along the ladder. For this reason, the early-versus-late rank contrast is not treated as the main inferential result. Instead, a permutation test of the association between over-estimation and task position gives $p < .001$. Even within the hard tail, over-estimation remains a minority outcome: 82\% of reported tasks from O52 to O60 were demonstrated.

The task results also provide a way to separate difficulty from point value. O45 carries the maximum 25 points but is deliberately easy; it was reported by 195 of the 203 students and produced no over-estimations. Thus, a high point value did not produce over-claiming when the task itself was easy. The other 25-point task, O60, is the hardest coding problem and has six over-estimations, but it was reported by only 99 students, so its low count cannot be interpreted as evidence of a weak effect of task value.

Across the hard tail, the highest over-estimation rates occur on tasks with low point values, rather than on the highest-valued tasks. Spearman correlation between over-estimation rate and ladder position is $\rho = +0.62$, but this should not be interpreted as evidence of a smooth monotone gradient: 51 of the 60 tasks have an over-estimation rate of zero, and ladder position is strongly associated with both task exposure ($r = -0.95$) and curricular time. The task-level evidence therefore supports a narrower conclusion. Self-report is highly accurate across the easy and middle portion of the ladder, while the observed over-estimation is concentrated at the hard end, where task difficulty rather than point value provides the more direct explanation.

\subsection{Discussion}
The findings suggest that self-report in a mastery-based dashboard is more useful than a simple question of whether students can be trusted to judge their own learning. At the cohort level, students' reports correspond closely with what they can demonstrate, but the remaining mismatch is structured rather than random. It is concentrated among a subset of students and emerges primarily when students reach the more difficult part of the task ladder. The practical implication is therefore not that self-report should be abandoned, but that it can be supplemented selectively where the likelihood of mismatch is greatest. For large engineering courses, this offers a possible middle ground between accepting every self-report without verification and attempting to verify every reported task.

The student-level pattern is particularly important for interpreting aggregate measures of self-assessment accuracy. A high overall correspondence can coexist with substantial differences in how individual students judge their own progress. Most students in this study did not contribute to the observed over-estimation, while a smaller group accounted for the discrepancies. This suggests that self-assessment accuracy should not be treated as a uniform property of the cohort. An average agreement measure can describe how well the system works overall while concealing the learners for whom the reporting mechanism is less reliable. For instructors, this means that the useful question is not simply whether students are generally accurate, but whether the system can identify situations in which a student's reported mastery deserves additional verification.

The task-level pattern provides a complementary explanation. The mismatch appears when students encounter tasks that require more complex integration of programming knowledge. This is consistent with the knowledge-in-pieces perspective, in which learners may possess productive fragments of understanding without yet being able to coordinate those fragments into a complete solution (diSessa, 1988, 1993). On simpler tasks, possessing the relevant fragment may be sufficient for successful performance and therefore provide a relatively clear basis for self-judgement. More demanding tasks require several pieces to be coordinated, creating a situation in which a student may recognise aspects of a solution and interpret that partial understanding as sufficient mastery. The observed gap between reporting and demonstration can therefore be understood as occurring near the boundary of a learner's demonstrated competence.

This interpretation should nevertheless be treated as one possible explanation rather than as a causal conclusion. The study measures the discrepancy between what students reported and what they demonstrated; it does not directly measure the cognitive processes that produced that discrepancy. In particular, the course used points, a public leaderboard, and graded task completion. A student who reported a difficult task that they could not subsequently demonstrate may therefore have been making a metacognitive judgement, taking a strategic risk, or combining both. The data cannot distinguish these possibilities. The appropriate interpretation of over-estimation in this study is consequently a behavioural one: the student's reported mastery was not confirmed by the eventual viva outcome. It should not be treated as direct evidence that the student lacked self-awareness.

The relationship with task value further supports the importance of considering task characteristics when interpreting self-report. A highly rewarded task did not necessarily produce more over-claiming when the task itself was accessible, whereas the more difficult tasks were where the mismatch became visible. This suggests that the challenge associated with completing a task may matter more than the incentive attached to claiming it. However, because task difficulty, task position, exposure, and curricular timing are closely related in the present ladder, these factors cannot be completely separated. The findings therefore support an association between task difficulty and over-estimation rather than establishing difficulty as its sole cause.

These findings have a direct implication for the design of mastery dashboards. Verification does not necessarily have to be distributed uniformly across all student-task claims. A more efficient design could use the dashboard to identify points in the learning sequence where self-reports are more likely to require confirmation and concentrate instructor or teaching-assistant verification there. Importantly, the present data suggest that the more dependable prediction is associated with the task rather than with a particular type of student. A system could therefore use task characteristics as a trigger for additional verification without first labelling individual learners as inaccurate. Such selective verification would preserve the scalability advantage of self-report while adding an independent check where the reporting mechanism is most vulnerable.

The discrepancy between self-report and demonstration can also be used as formative information rather than only as an assessment-control mechanism. When a student reports mastery but cannot demonstrate the task, the disagreement identifies a specific point at which the student's understanding and performance diverge. Making that discrepancy visible to the learner can provide more actionable feedback than a single aggregate examination score because it identifies the particular competency that needs further attention. This aligns with work showing that effective feedback and self-assessment are most useful when they help learners understand the gap between their current performance and the desired level of performance (Hattie \& Timperley, 2007; Nicol \& Macfarlane-Dick, 2006; Panadero et al., 2017; Yan \& Brown, 2017). Clear criteria for what constitutes successful demonstration may further help students make more accurate judgements about demanding tasks (Sadler \& Good, 2006).

The findings also highlight a requirement for the verification mechanism itself. If mastery claims are to be used as a basis for feedback or progression, the criterion against which those claims are checked must be applied consistently. The examiner analysis in this study provides only limited assurance because examiner and student groups were not independently crossed. Nevertheless, it points to the importance of shared rubrics, common exemplars, and periodic moderation when viva-based verification is distributed across multiple examiners (Huxham et al., 2012). In a scalable mastery system, reliability cannot rest entirely on the dashboard or on the student's self-report; it also depends on the consistency of the verification process.

Taken together, the results point to a design principle for scalable mastery-based learning: self-report can serve as the primary mechanism for recording progress, while independent verification can be targeted to the situations in which self-report is most likely to diverge from demonstrated mastery. The contribution of the present study is not to show that students are generally inaccurate, nor to argue that every self-report requires examination. Rather, it shows how a task-level comparison can reveal where the small amount of remaining mismatch is concentrated and thereby provide a basis for designing a more selective verification and feedback mechanism.

\subsection{Limitations}
\label{sec:limits}
Several limitations bound these findings. The study covers one course, one academic year, and one institution, so the reported magnitudes are local and the findings should be generalised primarily at the level of the observed structure rather than the exact rates. The one-sided design captures over-estimation but not under-estimation, and the strong ceiling effect limits our ability to examine whether ability predicts over-estimation. Because only the final viva outcome was retained, the number of attempts required to demonstrate a task is unknown. The design also cannot distinguish metacognitive error from strategic reporting, given the points and public leaderboard, or separate examiner effects from branch effects because most assistants examined students within a single branch. Finally, the course provided additional support to struggling students, which may have contributed to the low overall error rate.

\section{Conclusion}
\label{sec:conc}
This study examined task-level self-assessment by comparing students' reported programming-task mastery with their viva-demonstrated mastery in a first-year engineering course. Self-report showed high correspondence with demonstrated mastery at the cohort level, but the remaining mismatch was not distributed uniformly. Over-estimation was concentrated among a small group of students and at the harder end of the task ladder, while task point value showed less direct association with the observed mismatch. The findings therefore suggest that aggregate self-assessment accuracy can conceal where and for whom self-report becomes less reliable.

The main contribution of the study is to show the value of measuring this mismatch at the level of individual tasks rather than through a single overall self-rating and examination score. For mastery-based courses, this makes self-report usable as a scalable first layer of progress reporting while allowing independent verification to be directed toward tasks where mismatch is more likely to occur. The observed self-report--viva gap can also provide task-specific feedback to students, linking assessment of mastery with opportunities for recalibration. Future work should examine whether the same pattern holds across courses and institutions and develop a general task-level framework for selective verification and feedback in mastery-based learning environments.

\section*{Acknowledgements}
The authors thank the VicharanaShala Lab for Education Design (VLED), Indian Institute of Technology Ropar, for supporting this research and for access to the de-identified course records. We also thank the teaching assistants who conducted the vivas and the students whose participation made this study possible.

The authors used a generative AI assistant to help restructure and edit drafts of the text and to format the manuscript. The authors designed the study, checked all analyses, results, and references, and take full responsibility for the content.

\section*{References}
\footnotesize
\setlength{\leftskip}{1.6em}\setlength{\parindent}{-1.6em}
Andrade, H.~L. (2019). A critical review of research on student self-assessment. \textit{Frontiers in Education, 4}, Article 87. \url{https://doi.org/10.3389/feduc.2019.00087}\par\smallskip
Bennedsen, J., \& Caspersen, M.~E. (2019). Failure rates in introductory programming: 12 years later. \textit{ACM Inroads, 10}(2), 30--36. \url{https://doi.org/10.1145/3324888}\par\smallskip
Bloom, B.~S. (1968). Learning for mastery. \textit{Evaluation Comment, 1}(2), 1--12. (ERIC No.\ ED053419)\par\smallskip
Bodily, R., \& Verbert, K. (2017). Review of research on student-facing learning analytics dashboards and educational recommender systems. \textit{IEEE Transactions on Learning Technologies, 10}(4), 405--418. \url{https://doi.org/10.1109/TLT.2017.2740172}\par\smallskip
Boud, D., Cohen, R., \& Sampson, J. (1999). Peer learning and assessment. \textit{Assessment \& Evaluation in Higher Education, 24}(4), 413--426. \url{https://doi.org/10.1080/0260293990240405}\par\smallskip
de Ayala, R.~J. (2009). \textit{The theory and practice of item response theory}. Guilford Press.\par\smallskip
Deterding, S., Dixon, D., Khaled, R., \& Nacke, L. (2011). From game design elements to gamefulness: Defining ``gamification.'' \textit{Proceedings of the 15th International Academic MindTrek Conference}, 9--15. \url{https://doi.org/10.1145/2181037.2181040}\par\smallskip
diSessa, A.~A. (1988). Knowledge in pieces. In G. Forman \& P.~B. Pufall (Eds.), \textit{Constructivism in the computer age} (pp.\ 49--70). Lawrence Erlbaum Associates.\par\smallskip
diSessa, A.~A. (1993). Toward an epistemology of physics. \textit{Cognition and Instruction, 10}(2--3), 105--225. \url{https://doi.org/10.1080/07370008.1985.9649008}\par\smallskip
Dunlosky, J., \& Rawson, K.~A. (2012). Overconfidence produces underachievement: Inaccurate self-evaluations undermine students' learning and retention. \textit{Learning and Instruction, 22}(4), 271--280. \url{https://doi.org/10.1016/j.learninstruc.2011.08.003}\par\smallskip
Ehrlinger, J., Johnson, K., Banner, M., Dunning, D., \& Kruger, J. (2008). Why the unskilled are unaware: Further explorations of (absent) self-insight among the incompetent. \textit{Organizational Behavior and Human Decision Processes, 105}(1), 98--121. \url{https://doi.org/10.1016/j.obhdp.2007.05.002}\par\smallskip
Embretson, S.~E., \& Reise, S.~P. (2000). \textit{Item response theory for psychologists}. Lawrence Erlbaum Associates.\par\smallskip
Falchikov, N., \& Boud, D. (1989). Student self-assessment in higher education: A meta-analysis. \textit{Review of Educational Research, 59}(4), 395--430. \url{https://doi.org/10.3102/00346543059004395}\par\smallskip
Flavell, J.~H. (1979). Metacognition and cognitive monitoring: A new area of cognitive-developmental inquiry. \textit{American Psychologist, 34}(10), 906--911. \url{https://doi.org/10.1037/0003-066X.34.10.906}\par\smallskip
Harden, R.~M. (2002). Learning outcomes and instructional objectives: Is there a difference? \textit{Medical Teacher, 24}(2), 151--155. \url{https://doi.org/10.1080/0142159022020687}\par\smallskip
Hattie, J., \& Timperley, H. (2007). The power of feedback. \textit{Review of Educational Research, 77}(1), 81--112. \url{https://doi.org/10.3102/003465430298487}\par\smallskip
Huxham, M., Campbell, F., \& Westwood, J. (2012). Oral versus written assessments: A test of student performance and attitudes. \textit{Assessment \& Evaluation in Higher Education, 37}(1), 125--136. \url{https://doi.org/10.1080/02602938.2010.515012}\par\smallskip
Joughin, G. (1998). Dimensions of oral assessment. \textit{Assessment \& Evaluation in Higher Education, 23}(4), 367--378. \url{https://doi.org/10.1080/0260293980230404}\par\smallskip
Keuning, H., Jeuring, J., \& Heeren, B. (2018). A systematic literature review of automated feedback generation for programming exercises. \textit{ACM Transactions on Computing Education, 19}(1), Article 3. \url{https://doi.org/10.1145/3231711}\par\smallskip
Koo, T.~K., \& Li, M.~Y. (2016). A guideline of selecting and reporting intraclass correlation coefficients for reliability research. \textit{Journal of Chiropractic Medicine, 15}(2), 155--163. \url{https://doi.org/10.1016/j.jcm.2016.02.012}\par\smallskip
Kruger, J., \& Dunning, D. (1999). Unskilled and unaware of it: How difficulties in recognizing one's own incompetence lead to inflated self-assessments. \textit{Journal of Personality and Social Psychology, 77}(6), 1121--1134. \url{https://doi.org/10.1037/0022-3514.77.6.1121}\par\smallskip
Kuder, G.~F., \& Richardson, M.~W. (1937). The theory of the estimation of test reliability. \textit{Psychometrika, 2}(3), 151--160. \url{https://doi.org/10.1007/BF02288391}\par\smallskip
Nelson, T.~O., \& Narens, L. (1990). Metamemory: A theoretical framework and new findings. In G.~H. Bower (Ed.), \textit{The psychology of learning and motivation} (Vol.\ 26, pp.\ 125--173). Academic Press. \url{https://doi.org/10.1016/S0079-7421(08)60053-5}\par\smallskip
Nicol, D.~J., \& Macfarlane-Dick, D. (2006). Formative assessment and self-regulated learning: A model and seven principles of good feedback practice. \textit{Studies in Higher Education, 31}(2), 199--218. \url{https://doi.org/10.1080/03075070600572090}\par\smallskip
Panadero, E., Jonsson, A., \& Botella, J. (2017). Effects of self-assessment on self-regulated learning and self-efficacy: Four meta-analyses. \textit{Educational Research Review, 22}, 74--98. \url{https://doi.org/10.1016/j.edurev.2017.08.004}\par\smallskip
Sadler, P.~M., \& Good, E. (2006). The impact of self- and peer-grading on student learning. \textit{Educational Assessment, 11}(1), 1--31. \url{https://doi.org/10.1207/s15326977ea1101_1}\par\smallskip
Shrout, P.~E., \& Fleiss, J.~L. (1979). Intraclass correlations: Uses in assessing rater reliability. \textit{Psychological Bulletin, 86}(2), 420--428. \url{https://doi.org/10.1037/0033-2909.86.2.420}\par\smallskip
Watson, C., \& Li, F.~W.~B. (2014). Failure rates in introductory programming revisited. \textit{Proceedings of the 2014 Conference on Innovation \& Technology in Computer Science Education}, 39--44. \url{https://doi.org/10.1145/2591708.2591749}\par\smallskip
Yan, Z., \& Brown, G.~T.~L. (2017). A cyclical self-assessment process: Towards a model of how students engage in self-assessment. \textit{Assessment \& Evaluation in Higher Education, 42}(8), 1247--1262. \url{https://doi.org/10.1080/02602938.2016.1260091}\par\smallskip
Zell, E., \& Krizan, Z. (2014). Do people have insight into their abilities? A metasynthesis. \textit{Perspectives on Psychological Science, 9}(2), 111--125. \url{https://doi.org/10.1177/1745691613518075}\par

\end{document}